\documentclass[1p]{elsarticle}
\usepackage[utf8]{inputenc}
\usepackage{rotating}
\usepackage{makecell}
\usepackage{hyperref}
\usepackage[title]{appendix}
\hypersetup{
    colorlinks=true,
    linkcolor=blue,
    filecolor=magenta,      
    urlcolor=cyan,
}
\usepackage[english]{babel}
\usepackage{etoolbox}
\usepackage{amssymb}
\usepackage{booktabs}
\usepackage{multirow}
\renewcommand{\href}[2]{#2}

\newcommand{\EX}[1]{\textcolor{blue}{#1}}
\renewcommand{\EX}[1]{#1}

\journal{Computer Communications VSI: SI Marco Ajmone}

\begin{document}

\begin{frontmatter}

%\title{Neural language models for network configuration verification, synthesis and cross-Vendor translation: opportunities and reality check}
\title{Neural language models for network configuration:\\Opportunities and reality check}

\author[inst1]{Zied Ben Houidi}
\affiliation[inst1]{organization={Huawei Technologies France SASU},%Department and Organization
            addressline={20 quai du point du jour}, 
            city={Boulogne-Billancourt},
            %postcode={01234}, 
            country={France}}
\author[inst1]{Dario Rossi}
%\date{November 2021}
%\makeatletter
% \def\@copyrightspace{\relax}
% \makeatother
%  \setcopyright{none} 

\begin{abstract}
Boosted by deep learning, \emph{natural language processing} (NLP) techniques have recently seen spectacular progress, mainly fueled by breakthroughs both in representation learning with word embeddings (e.g. word2vec) as well as novel architectures (e.g. transformers). This success quickly invited researchers to explore the use of NLP techniques to  other field, such as \emph{computer programming languages}, with the promise to automate tasks in software programming (bug detection, code synthesis, code repair, cross language translation etc.). By extension, NLP has potential for application to network configuration languages
as well, for instance considering tasks such as network configuration verification, synthesis, and cross-vendor translation. In this paper, we survey recent advances in deep learning applied to programming languages, for the purpose of code verification, synthesis and translation: in particularly, we review their training requirements and expected performance, and qualitatively assess whether similar techniques can benefit corresponding use-cases in networking.
 \end{abstract}

%%Graphical abstract
% \begin{graphicalabstract}
% \includegraphics{grabs}
% \end{graphicalabstract}

%%Research highlights
% \begin{highlights}
% \item Research highlight 1
% \item Research highlight 2
% \end{highlights}

\begin{keyword}
%% keywords here, in the form: keyword \sep keyword
Natural language processing \sep Code verification, synthesis, translation \sep Network configuration verification, synthesis translation
%% PACS codes here, in the form: \PACS code \sep code
%\PACS 0000 \sep 1111
%% MSC codes here, in the form: \MSC code \sep code
%% or \MSC[2008] code \sep code (2000 is the default)
%\MSC 0000 \sep 1111
\end{keyword}
\end{frontmatter}

%\maketitle

\section{Introduction}
Network operators often rely on a heterogeneous set of equipments from different vendors, each of which uses different proprietary configuration languages. Such heterogeneity poses a number of challenges, that have the potential to turn the dream of intent-based, fully autonomous and self-driving networks into a waking nightmare.
While the reliance on multiple vendors is a very logical choice from a business perspective, it makes network management a quite complex task. Efficiently managing real networks needs  knowledge about the specifics of each of these multiple vendors, a skill that is rare in practice. This has pervasive consequences and affects many tasks from provisioning (i.e. generating new configurations) to verification, to monitoring, debugging and troubleshooting.

%and fueled by the wide adoption of the SDN paradigm, 
To counter this problem, the network community attempted to create additional abstractions to unify network control, of which Batfish\cite{fogel2015general} from Intentionet is one popular example. Batfish uses expert rules to transform each proprietary configuration into a unified language model, that can be used to verify the correctness of a configuration or visualize its effects before it is actually put in place.   
Similarly, many other approaches (see Sec.~\ref{sec:formal}) leverage formal methods to automate various aspects of network configuration, often transforming a network-wide configuration into a large logical formula, then for example adding specification constraints on top, to verify that the specifications match the configuration.
Although promising, such approaches still need humans in the loop, either to build or update the unified language model, or to manually design the numerous verification tests required for the new unified language. Devil's advocating, one can say that approaches like Batfish solve the problem of ``too many proprietary languages'' by adding another proprietary language~\cite{xkcd927}. 

At the same time, advances in deep learning based Natural Language Processing (NLP) techniques have opened new opportunities, allowing to perform tasks that seemed impossible a decade ago. For example, recent progress in unsupervised neural machine translation makes it possible today to translate between different languages, using mostly mono-lingual\footnote{Otherwise stated, neural language models can learn to translate from reading independently books in each language, as opposed to requiring a curated parallel corpora of books translated in many languages} corpora~\cite{lample2017unsupervised}, which are usually abundant. 
Text generation has also reached impressive performance~\cite{brown2020language}, as widely popularized by OpenAI GPT-3~\cite{gpt3}. This giant step in representation learning from natural language corpora naturally raises the question of whether these breakthroughs can profit other artificial languages, such as those used for computer programming or network configuration. 

Whereas neural NLP techniques have so far enjoyed a  limited adoption in the network community,  the last few years have witnessed  an emerging trend in the application of NLP technologies to  programming languages, with growing success on tasks such as code (i) verification, (ii) synthesis and even (iii) translation. 
We observe that the above three axes are very much in line  
with important questions regarding the benefits of NLP for network management, specifically:  (i) whether properly modeled  network configuration languages could ease the detection of misconfigurations or underspecifications, as it is possible to detect grammatical mistakes in natural language; (ii) whether recent advances in NLP could allow to make advances in router configuration synthesis, in a way that is similar to natural text and code generation; (iii)  whether it is feasible to automatically translate between routing configuration languages of different vendors -- which, if successful, could be a significant step towards reducing the interoperability gap.

In this paper we argue that, to some extent, progress on NLP for programming languages can be informative so as to estimate the potential of NLP for solving similar tasks on network configuration languages.
Yet, the state of progress in the programming language domain is not fully clear: generally speaking, it is difficult to discriminate hype from small-step methodological contributions, so to select which approach  among the numerous proposals could the most promising for network configuration. Second, it is also non-trivial to assess at which cost in terms of data and computing power such prowesses were made possible, and thus at which upfront cost it would be possible to apply NLP advances to network configuration use-cases.  Finally, it is even harder to estimate if, and  within which time-frame, such pioneering approaches can transition to useful products extensively used in real systems. 

The goal of this paper is  therefore to systematically analyze the progress of recent NLP technologies for artificial programming languages,  and assess their potential for application to the network configuration languages for management purposes.  In the remainder of the paper, we first overview state of the art in the  network management use-cases of configuration verification, synthesis and cross-vendor translation (Sec.~\ref{sec:usecase}). We then review  the recent literature on NLP-empowered programming languages for the use-cases of code verification, synthesis and cross-language translation  (Sec.~\ref{sec:progress}). We next make an explicit parallel between these two fields: \EX{considering both a broad view, as well as using an illustrative networking example}, we assess limits of current methods,  data and model training requirements, need for pre-processing, task complexity and expected performance  (Sec.~\ref{sec:transfer}). Based on this analysis, we  gather conclusive remarks and sketch a high-level research roadmap (Sec.~\ref{sec:conclusion}).

\section{NLP for ``Network Language'' Processing}\label{sec:usecase}
%%%%%%%%%%%%%%%%%%%%%%%%%%%%%%%%%%%%%%%%%%%%%%%%%%%%%%%%%%%%%%%%%%%%%%%%%%%%%%%

We first identify opportunities where NLP could help network operations. In this paper, we focus  on  three relevant use-cases of network configuration, introduced next, and overview how they are dealt with in the networking literature.
\EX{For the sake of illustration, throughout the paper we will also systematically make reference to the toy case of an ISP network administrator that has to generate, verify or translate the BGP configurations of one or all its routers}.

%The most straightforward and esiest is to find reachability-related violations. 
%BGP could learn the underlying forwarding behavior, translate it into intent and compare the forwarding behavior with the intended one. One big challenge there (shared also with traditional rule-based approaches) is how to represent user or customer intent and requirements. 

%Generating code and letting humans verifying it is cumbersome as it is more difficult to learning. Similarly to what was proposed by Roziere \textit{et al.}\cite{roziere2021leveraging}, one counter measure is to rely on automated unit-tests .

\subsection{NLP Potential for Network Configuration}
\paragraph{Configuration Verification}%%%%%%%%%%%%%%%%%%%%%%%%%%%%%%%%%%%%%%%%%%%%%%%%%%%%%%%%%%%%%%%%%%%%%%%%%%%%%%%
A critical task receiving growing attention in the networking community is configuration verification, with the accessory goals of detecting and fixing anomalies in the configuration. The classic examples are reachability (e.g., how can we verify that a network configuration is free from e.g. black holes, loops, or other conflicting rules) and compliance (e.g., verifying that the configuration satisfies a given policy).
%%%%
As we shall overview later, existing approaches rely on formal methods heavily involving human experts. The question (and opportunity) here is whether data-driven NLP  methods can be of any help. For example, an open question is whether  language models (trained in an unsupervised manner on mostly correct network-wide configurations) coupled with  classic anomaly detection techniques can spot  misconfigurations or inconsistencies.
A related question is whether the further addition of a  supervised layer (fine-tuned on certainly correct  configurations) can assist in repairing the errors.
\EX{If this task could be successfully implemented by NLP then, for the BGP toy case, an operator would  only have to submit the configuration files of its network and inspect an output report about syntax and reachability errors. If the operator further provided specifications about routing policies, the checker would verify if routes are not redistributed in ways that violate these policies.}

% IF YWE DON"T TALK ABOUT IT L:TER,
%A closely related problem is the existence of ``phantom'' outdated commands that are still on routers but that are not used in practice. Can NLP tools help spotting such useless commands?  

\paragraph{Configuration  Synthesis}%%%%%%%%%%%%%%%%%%%%%%%%%%%%%%%%%%%%%%%%%%%%%%%%%%%%%%%%%%%%%%%%%%%%%%%%%%%%%%%
One recurrent task in networking is the configuration of various devices, either to satisfy customer requests (i.e. provisioning) or to later optimize various resources when adapting to network conditions. The input would be customer requests or SLAs expressed either in some arbitrary specification language (possibly complemented by natural language description in the longer term) and the output is a set of configurations that satisfy this goal. The paramount  question  in this case is to what extent can NLP tools help automating such generative process -- which is inline with the long-term wish of intent-based networking.  

\EX{With reference to the BGP toy case, if this task is successfully implemented by NLP, the network operator would only have to master a high-level specification language, describing reachability requirements depending on negotiated policies, and the NLP agent would automatically generate the corresponding configurations.}

\paragraph{Configuration Translation}%%%%%%%%%%%%%%%%%%%%%%%%%%%%%%%%%%%%%%%%%%%%%%%%%%%%%%%%%%%%%%%%%%%%%%%%%%%%%%%
Finally, as earlier mentioned, another significant problem faced by large operators is due to their control of large heterogeneous fleets of network devices from multiple vendors, each using own proprietary languages. Besides the day-to-day management hurdle, migrating an equipment from one vendor to the other today can be cumbersome, not only in terms of translation but also in terms of network configuration understanding and cleaning.
%%%
Therefore, the potential for NLP is manifest, as it would be desirable to leverage neural language techniques for learning to automatically translate between configuration languages of various vendors.  If successful, this would not only ease the advent of the self driving network vision (one language to control them all) but also the management of legacy networks. 

\EX{With reference to  the BGP toy case, a network operator who needed to replace a router with another from a different vendor, would only have to submit the legacy configuration lines to obtain the new one in the target language.}

\subsection{Current State of the Art in Network Configuration}\label{sec:formal}
At the same time, existing attempts to automate network configuration verification and synthesis are mostly rule-based, often relying on formal methods, and have not yet exploited NLP techniques to the best of our knowledge. %\DR{minor exceptions we want to point out?}\ZH{not found}
 
Outside the above network configuration use cases, we point out that recent NLP techniques are used in networking, such as 
word embeddings to learn representations\cite{CoNEXT21-b,ring2017ip2vec,cohen2020dante} and transformers applied to graphs\cite{velickovic2018graph,kool2018attention}.
%, but they generally fall outside the scope of network configuration use-cases and is thus orthogonal with respect to our focus. 
In this section, focusing on network configuration, and without aiming at presenting an exhaustive survey of related literature, we compactly present in Tab.\ref{tab:sota_network} a simple taxonomy of the state of the art, along with representative samples for each category.

% \subsection{Representation learning and beyond}
% \ZH{self-cite mainly, so either extend or move to conclusion as: there are other applications of NLP in networking}
%Finally, beyond network configuration use cases discussed in this paper, several networking problems can benefit from advances in NLP. First, several representation learning tools built initially for NLP can be applied to analyze and reason about network data, of which a large portion is categorical in nature and can be conceptually seen as words in a language~\cite{}.
% We did similar work in the past in FRC, applying word embeddings on domain names~\cite{PCT/CN2020/102259} for clickstream identification, on AP names~\cite{PCT/EP2021/057178} for movement prediction and recently on IP addresses for Darknet clustering~\cite{CoNEXT21-b}. Similar work in the community is IP2vec~\cite{ring2017ip2vec} and Dante~\cite{cohen2020dante}.
%Beyond representation learning with word embeddings, transformers~\cite{NIPS2017_7181} are being used to represent nodes and edges in graph network problems\cite{velickovic2018graph,kool2018attention}. %We have been also exploring the use of chat-bot inspired technologies to learn how to ``speak'' protocols to automate the operation of honeypots~\cite{PCT/EP2021/076753}. Such NLP inspired applications are beyond the scope of this report. So is the case for other applications of NLP to process natural language involved in networking such as RFCs, trouble tickets etc.  

\begin{table*}[t]
    \centering
    \caption{Simple taxonomy and non-exhaustive list of example work related to network management and configuration}
    \begin{tabular}{lp{5cm}p{5cm}}
    \toprule
    Use-case    & Methods & Network application \\
    \midrule
    Verification    &  \begin{tabular}{@{}l@{}}
											Model checking~\cite{prabhu2020plankton}\\
											SMT~\cite{weitz2016scalable,beckett2017general,tian2019safely}\\
											Graph-based~\cite{gember2016fast}\\
											Custom~\cite{feldmann1999netdb,fogel2015general,steffen2020probabilistic,abhashkumar2020tiramisu}\\
											\end{tabular} 
                    &  \begin{tabular}{@{}l@{}}
											BGP-only\cite{weitz2016scalable}\\ 
											ACL~\cite{tian2019safely}\\
											Subset of protocols~\cite{steffen2020probabilistic,gember2016fast,prabhu2020plankton}\\ 
											General~\cite{fogel2015general,beckett2017general,abhashkumar2020tiramisu}
										 \end{tabular} 
                    %&  
										\\
    \midrule
    \begin{tabular}{@{}l@{}}
    Synthesis \\ 
    % (generation or update)
    \end{tabular} 
                & 
                \begin{tabular}{@{}l@{}}
                SMT~\cite{beckett2017network,tian2019safely}\\
                Stratified Datalog~\cite{el2018netcomplete,el2017network}\\
                Custom model~\cite{liu2018automatic}
                % (generation or update)
                \end{tabular} 
                 &  \begin{tabular}{@{}l@{}}
                 BGP-only~\cite{beckett2017network}\\
                 ACL\cite{tian2019safely} \\
                 BGP/OSPF/Static~\cite{el2018netcomplete,el2017network}\\
                 General\cite{liu2018automatic}  
                 \end{tabular} 
                % &  
								\\ 
    \midrule
    Explanation     & \begin{tabular}{@{}l@{}}
										Custom~\cite{birkner2018net2text,birkner2020config2spec}
                    \end{tabular} 
                    & \begin{tabular}{@{}l@{}} 
										OSPF/BGP~\cite{birkner2018net2text} \\
                    Forwarding state~\cite{birkner2018net2text} \\
                    \end{tabular} 
								\\
                    
    \bottomrule
    \end{tabular}
    \label{tab:sota_network}
%     \DR{
% Otherwise, thinking loud. The axes are:
% Method: SMT solver, symbolic partioning, model checking
% Application: BGP or ACL
% Use case: conf verification (spec=?conf), generation or update (genconf), explanation ()
% }
\end{table*}

\paragraph{Configuration Verification}
As summarized in Table~\ref{tab:sota_network}, verification methodologies currently used 
for network configuration  are not leveraging NLP yet and can be mapped into several families, namely: (i) graph models or (ii) custom graph-based abstractions and exploration, (iii) explicit state model checking on top of a formal language and (iv) satisfiability modulo theory.
In more details, Netdb~\cite{feldmann1999netdb} was among the first
seminal attempts to automate the parsing, modelling, and the correctness verification of network wide configuration files, which received more attention lately.  Batfish~\cite{fogel2015general} is one popular example which builds a data plane model from router configurations and uses it among others for configuration verification.
\href{https://dl.acm.org/doi/pdf/10.1145/2934872.2934876}{ARC}\cite{gember2016fast} builds a graph-based abstraction of the control plane from network configuration files and uses it to analyze the control plane under arbitrary failures, without generating the data plane. 
\href{https://dl.acm.org/doi/abs/10.1145/2983990.2984012}{Bagpipe}~\cite{weitz2016scalable} uses an SMT solver to verify that BGP configurations satisfy the policies expressed by the operator. Minesweeper\cite{beckett2017general} also transforms network configuration files into a logical formula that captures the final state or behavior of the data plane, that is then combined with intended properties or desired specifications to see if both can match. 
\href{https://www.usenix.org/conference/nsdi20/presentation/prabhu}{Plankton}\cite{prabhu2020plankton} uses explicit-state model checking (together with symbolic partioning) to considerably speed up network policies verification.
%Other work (e.g.~\cite{abhashkumar2020tiramisu,}) proposed network verification tools
\href{https://dl.acm.org/doi/10.1145/3387514.3405900}{NetDice}~\cite{steffen2020probabilistic} analyzes the probabilities that certain properties hold in the network without fully enumerating all possible link failures. NetDice input is, however, already curated network configurations expressing BGP, OSPF, ECMP, and static route properties and, methodology-wise, the paper heavily relies on domain knowledge. 
Finally, instead of relying on too general search strategies used by SMT solvers during verification, \href{https://www.usenix.org/system/files/nsdi20-paper-abhashkumar.pdf}{Tiramisu}~\cite{abhashkumar2020tiramisu} builds its own network model and search strategies that are more efficient for network models: for example, it leverages routing algebra to verify paths in the graph in one shot (instead of emulating protocols as done with explicit-state model checking).

\EX{With reference to the BGP toy case, the network operator currently has to master one of the declarative languages to express its BGP routing policies or intents, and use any of the existing tools above to evaluate whether the configurations match these policies. A challenge there is to actually derive the properties that must be satisfied. For example, Bagpipe~\cite{weitz2016scalable} infers such desired policies from real AS configurations or from prior research work (e.g., the famous Gao-Rexford conditions~\cite{gao2001stable}). Unlike such exact methods, applying unsupervised anomaly detection methods on learned NLP representations could be promising to detect and locate anomalies in configurations and policies without incurring the burden of having to  formally express them. Alternatively, if the specifications or intents of the operator are available, then NLP methods could broaden the current set of exact methods, e.g, by learning to generalize over a larger set of specification languages.}

\paragraph{Configuration Synthesis}
Automatic generation of network configuration is another active area where, e.g., \href{https://dl.acm.org/doi/abs/10.1145/3341302.3342088}{Jinjing}\cite{tian2019safely} helps Alibaba's 
operators to transform their declarative configuration intents into ACL rule configuration updates.
Alibaba group has also developed \href{https://dl.acm.org/doi/abs/10.1145/3229584.3229585}{NetCraft}~\cite{liu2018automatic}, a tool that builds a unified network model and uses it to  safely manage the life cycle of network configuration. 
Prior to that, \href{https://dl.acm.org/doi/abs/10.1145/3062341.3062367}{Propane/AT}~\cite{beckett2017network} builds new abstractions to synthesise BGP configurations with correctness proofs from high-level specifications and requirements. \href{https://arxiv.org/pdf/1611.02537.pdf}{SyNet}~\cite{el2017network} augments stratified datalog to perform synthesis for protocols that can be expressed in this language, leading the authors to support OSPF, Static routes and a simplified version of BGP.  \href{https://netcomplete.ethz.ch/files/netcomplete-nsdi2018.pdf}{Netcomplete}\cite{el2018netcomplete} later complemented the work with full support of BGP and orders of magnitude faster computation. 
Finally, thinking about synthesis in a broader sense, it is worth mentioning that similar formal methods have been used to programmatically (i) generate protocol code (but not configurations) from RFC requirements\cite{yen2021semi}, and (ii) specifications from lower-level axiomatic requirements~\cite{houidi2016knowledge}. 

\EX{With our BGP example in mind, similarly to the verification use case, the operator has only to master a high-level specification language, more concise and thus easier to express than actual configuration language. While the configuration is currently generated with algorithms (such as SMT, datalog or other custom models) that do not leverage NLP, in this paper we consider the question of to what extent NLP advances could help the network expert in his operational tasks.}

\paragraph{Configuration Explanation}
Whereas translation of configuration across languages has not been heavily investigated so far, a related task consists in summarizing and explaining configurations to the human operators.
%%%
In this context, a first body of literature empirically studied the evolution of network configurations in the wild~\cite{benson2009unraveling,kim2011evolution,lee2008automate}.  For example, using custom parsers, authors perform a longitudinal evaluation over a time period of two~\cite{lee2008automate} to five~\cite{kim2011evolution} years, confirming that network configuration grow more complex over time.
A second body of literature attempted to automate the extraction of human-readable insights from network configuration, or indirectly from its forwarding state.
For the latter, Net2text~\cite{birkner2018net2text} aims at generating highly interpretable text explaining network forwarding behavior.
\href{https://www.usenix.org/conference/nsdi20/presentation/birkner}{Config2spec}~\cite{birkner2020config2spec} transforms router configurations into formal specifications, using  Batfish\cite{fogel2015general} to parse router configurations -- which helps operators in formalizing policies, but could be also a first step towards building datasets for more systematic usage of NLP techniques. \EX{In the BGP toy case and beyond, generative abilities of NLP techniques popularized in the latest years by, e.g., GPT-3, are promising to further provide comments and explanations in a native human language.}

% \href{https://dl.acm.org/doi/abs/10.1145/2068816.2068863}{Kim \textit{et al.}}~\cite{kim2011evolution} (and \href{https://ieeexplore.ieee.org/abstract/document/4534108}{Lee \textit{et al.}}~\cite{lee2008automate})  studied 5 (resp. 2) years of network configuration files focusing on the evolution of configurations and changes. Using custom parsers, they confirm that network configuration grow more complex over time. 

%of spec to conf to train machine learning models. Config2spec uses batfish to parse router configurations.

%\subsubsection{Leftover \DR{either we dispatch the following  references into the relevant section, or we delete them}}
%Many other recent tools adopt similar formal approaches for network configuration understanding, verification and automation~\cite{abhashkumar2020tiramisu,birkner2018net2text,steffen2020probabilistic,el2017network,el2018netcomplete,rivera2020polanco}, \DR{either you split them in categories per-application or per-methodology,  or we remove them} reviewing them is beyond the scope of the current paper. They are telling about the need to automate network configuration operations.

\section{NLP for ``Computer Language'' Processing}
Whereas spectacular progress~\cite{gpt3} in NLP has primarily benefited ``natural'' language applications,  a  research trend emerged lately for application of NLP to ``artificial'' programming languages.  In particular, recent NLP applications fall within the areas of code verification, code synthesis and code translation -- which mirror the use cases above that we envision for networking, and that we overview in this section.

\subsection{Code Verification}%%%%%%%%%%%%%%%%%%%%%%%%%%%%%%%%%%%%%%%%%%%%%%%%%%%%%%%%%%%%%%%%%%%%%%%%%%%%%%%
A considerable recent effort has explored the use of neural networks and NLP for bug detection and code verification.
\href{https://arxiv.org/pdf/1906.00307.pdf}{Habib \textit{et al.}}\cite{habib2019neural} studied the opportunity of formulating bug-finding as a classification problem trained on examples of buggy and bug-free code snippets. Richter and Wehrheim~\cite{richter2021deepmutants} stress  that existing neural bug detection techniques  work only for unrealistically generated bugs, and proposed novel ways to build realistic training datasets. This joins the conclusions of \cite{habib2019neural}, in that despite having good quantitative performance, the models struggled to understand obvious program properties.

As often the case in machine learning, learning good latent representations is key to good performance in later tasks~\cite{tian2020rethinking}: in the software context, code2vec~\cite{alon2019code2vec} is a notable example. Code2vec first transforms the code into an Abstract Syntax tree (AST) then builds embeddings leveraging the multiple AST paths between program entities. Code2vec learns the representation of each path together with the representation of an aggregation of paths. As a use case, Code2vec allows to successfully predict the name of a method from the vector representation of its body. Also, similarly to word2vec, code2vec learns method name vectors that interestingly capture semantic similarities between code snippets.
Encouraged by this success, \href{https://arxiv.org/pdf/1911.12863.pdf}{Briem~\textit{et al.}}~\cite{briem2019using} confirmed the potential of code2vec representation  in other use cases than function naming, as e.g., to find simple bugs such off-by-one bugs in Java.
%%%
To further enhance code representation learning for bug-detection, \href{https://dl.acm.org/doi/10.1145/3360588}{Li~\textit{et al.}}~\cite{li2019improving} complement the local context extracted from the generated AST (used by code2vec) with a global context coming from Program Dependence Graph (PDG) and Data Flow Graph (DFG): the latter allows to take into account also the ``far-apart'' dependencies between the various methods used in the code. For the local context part, they apply word2vec~\cite{mikolov2013efficient} on the sequence of nodes in AST paths and use  node2vec~\cite{grover2016node2vec} for the global context given by PDG  and DFG graphs. 

\textit{Overall, we see that significant progress in code verification is achieved by leveraging the existing domain expert knowledge to increase the power of the learned NLP representations -- which  can be expected to hold for other artificial languages.}.

\subsection{Code Synthesis}\label{sec:progress} 
The goal of  code synthesis  is, starting from a text description in natural language or sometimes simply a function name, to generate the right code snippet.
Machine learning on the other hand promises to get rid of these manual efforts: for example, huge amounts of commented code are available online,  so that one can learn to transform the comments describing a function into the actual code of the function.

One notorious example of such attempts is represented by the 2020 NLC2CMD~\cite{agarwal2021neurips} NeurIPS competition on learning how to transform natural language descriptions into CLI bash syntax: since the task takes natural language as input,  an appealing choice is to start from pre-trained large NLP models (such as GPT-3) and only refine the training for the task at hand, an approach followed by several teams of 2020 NLC2CMD and by subsequent work~\cite{ahmad2021unified, chen2021evaluating,austin2021program}.

\href{https://arxiv.org/pdf/2103.06333.pdf}{PLBART}\cite{ahmad2021unified} is 
a bidirectional and autoregressive transformer for program and language understanding and generation, that achieves state of the art performance in code summarization, synthesis and translation.
As usual practice with transformers for NLP, PLBART is pre-trained on huge amounts of unlabeled data  (see later Sec.\ref{sec:transfer:data} for details), of both programming (GitHub) and natural (StackOverflow) languages.   Overall, the authors stressed the importance of task-independent pre-training (to achieve a good level of program and language understanding) before specializing in various tasks.  To illustrate its high representational power, authors show that thanks to pre-training,  PLBART learns deep aspects of code such as syntax, naming conventions etc. 

%PLBART }\cite{ahmad2021unified}on code summarization, generation, translation, program repair, clone detection, and vulnerability detection tasks
%This represents respectively 36.4 B, 28B and 6.7B tokens.

%%%%% Codex\cite{chen2021evaluating} Codex was obtained by fine-tuning GPT-3 models containing up to 12B parameters. It was trained on 179GB of unique python files scraped from GitHub, finetuning from GPT-3 family models that have already good natural language processing capabilities. 

%(from 244M to 137B parameters) \cite{austin2021program} 

\href{https://arxiv.org/pdf/2107.03374.pdf}{Open AI Codex}\cite{chen2021evaluating} is a related effort, whose production version empowers the \href{https://copilot.github.com/}{GitHub Copilot project} assisting developers with code autocompletion from docstring-like descriptions and function names.  Codex is obtained by fine-tuning GPT-3 models over python sources from GitHub, and  exhibit significant improvements over the GPT-3 performance trained on natural language only:  if given the chance to produce many code attempts, it can solve all the problems in the \href{https://github.com/openai/human-eval}{Human-eval} dataset.  At the same time, limits remain since Codex (i)  is not sample-efficient, i.e. humans learn to code from significantly fewer examples,  (ii)  can invoke undefined functions and variables and struggles with long/high-level docstring and (iii) has trouble binding variables to objects. 
Along the same direction, \href{https://arxiv.org/pdf/2108.07732.pdf}{Austin~\textit{et al.}}\cite{austin2021program} performed a critical and thorough evaluation of the abilities of large
NLP models to  synthesize code, shedding light on how, e.g., performance varies according to model size,  how chat interaction with a human in natural language can reduce the error, etc.

%{Program synthesis with large language models} 
%neural language models to synthesize code. Their analysis sheds light on how performance varies according to model size, how chat interaction with a human in natural language can reduce the error, and where and why current models fail to perform the task.

Finally, we point out that relevant work targeting the synthesis of network-related software  (but not network-related configurations) recently started to appear, with e.g., NLP techniques applied to text of IETF Request For Comments (RFC) normative documents for the sake of either auto-discovering and fixing ambiguities \cite{yen2021tools} or automatically generating  protocol implementations \cite{yen2021semi}. 

{\it Overall, we see that the success of code synthesis is certainly tributary to learning good representations from raw code. Equally importantly however, it is the availability of docstrings and code comments that allowed to fine tune large language models to perform synthesis. Hence, the value of commenting code extends beyond the realm of human understanding -- whenever available, natural language comments accompanying code can complement the formal representation of artificial language.}

\subsection{Code Translation} %%%%%%%%%%%%%%%%%%%%%%%%%%%%%%%%%%%%%%%%%%%%%%%%%%%%%%%%%%%%%%%%%%%%%%%%%%%%%%%

Driven by the success of deep learning in NLP translation from one human language to another, 
recent work tackled the equivalent problem of code translation from one programming language to another.

\href{https://arxiv.org/pdf/1802.03691.pdf}{Chen~\textit{et al.}}~\cite{chen2018tree} is among the first attempts at using recurrent neural networks (RNN) for programming language code translation.
They first observe that, starting from a certain length, RNNs struggle to generate syntactically correct programs. Inspired by traditional approaches where the human translator builds a rule-based correspondence between the grammars of the two programming languages, they propose to exploit the modularity of the translation task, transforming it into a translation between a source (parsed) tree and a target (parsed) tree.

%. Their results show superiority of the neural network approach, including on real world projects.

Inspired by unsupervised translation for natural languages~\cite{lample2017unsupervised}, more recent work~\cite{roziere2020unsupervised,roziere2021leveraging,roziere2021dobf} leverages transformer architectures for code translation. Similarly to unsupervised translation in natural language, the goal is to find a latent representation that is common between the two languages, such that sentences or code snippets with the same meaning have the same representation in the latent space. 
\href{https://arxiv.org/abs/2006.03511}{La Chaux~\textit{et al.}}~\cite{roziere2020unsupervised} proposed the first of such seq2seq architectures,  leveraging attention and encoder/decoder blocks.
The first step is to pre-train a cross-lingual language model, which can be done in a fully unsupervised manner from an independent corpus, especially when few anchor points exist between the two languages (e.g. \texttt{for}, \texttt{while} and other keywords for source code):  after training, the encoder can turn any input sequence into a common latent representation vector.
The second step  is to learn a decoder that can generate a sentence in the target domain: simplifying, this task is implemented using a denoising autoencoder, trained to generate a sentence from a noisy version of it.  The above architecture is shown~\cite{roziere2020unsupervised}  to outperform all available commercial rule-based software, on a custom evaluation dataset built by leveraging the GeeksforGeeks online platform\footnote{https://practice.geeksforgeeks.org/}.

%In practice, a third refinement task helps achieving even better results: using supervised learning to back-translate from the translated sentence or snippet to the original one again.
%Our method relies exclusively on monolingual source code, requires no expertise in the source or target languages, and can easily be generalized to other programming languages. 

In a subsequent work, \href{https://arxiv.org/abs/2110.06773}{Roziere \textit{et al.}}\cite{roziere2021leveraging} acknowledge one major limitation of their previous natural language methods: namely the use of back-translation which implies training on possibly noisy inputs and learning to generate noisy ones. While a small noise in natural language may be imperceptible, this can lead to an incorrect program in more structured source code. To counter this, the authors simply propose to leverage automated unit-testing to filter incorrectly generated code.

{\it Overall, we see that NLP is able to extract powerful representations from raw code that,  by leveraging multiple independent corpora each representing a different programming language, are also able to partly address automated translation among languages.}

%%%%%%%%%%%%%%%%%%%%%%%%%%%%%%%%%%%%%%%%%%%%%%%%%
\subsection{Other software-related tasks} %%%%%%%%%%%%%%%%%%%%%%%%%%%%%%%%%%%%%%%%%%%%%%%%%%%%%%%%%%%%%%%%%%%%%%%%%%%%%%%
In addition, other software related tasks are currently being investigated in the literature, that we briefly overview next as they  potentially open other opportunities for network configuration purposes. While we disregard them in this paper due to lack of space, they are at least worth mentioning.

\paragraph{Automated code documentation}
\href{https://arxiv.org/ftp/arxiv/papers/2104/2104.02443.pdf}{CodeTrans}~\cite{elnaggar2021codetrans} is an example of a pre-trained model to perform a variety of such tasks such as (i) code documentation Generation, (ii) source code summarization (iii) code comment generation, (iv) commit message generation, (v) API sequence recommendation. 
One question is whether similar methods could be helpful for network configuration explanation  (for example, to document thousand-lines long configuration files of complex routing policies). 
\EX{Thinking back about our network administrator, automatically generating BGP policy summarizations or explanations can speed up the reasoning, easing the updates of existing policies as well as the transfer of duties between various administrators.}

\paragraph{Automated code completion/repair}

Code completion~\cite{pradel2020neural}  and repair~\cite{yasunaga2021break} could also find applications in networking: network  configuration completion is tackled e.g.,  in~\cite{el2018netcomplete}, by however using formal methods and not machine learning.
In automated code repair, the task is to transform a program with syntax errors into a correct one -- a use-case that sits in between code verification (as code is altered) and synthesis (as the code generation is more limited). Code repair is tackled for example in  \href{https://arxiv.org/abs/2106.06600}{Break-It-Fix-It (BIFI)}~\cite{yasunaga2021break} by a sophisticated yet more realistic training approach.
Instead of creating pairs of (bad, good) programs for training using heuristics, which leads to non representative training data and models that overfit to unrealistic synthetic errors, BIFI introduces a breaker network that is trained to generate realistic errors, in addition to a critic that checks the Fixer's output and augment training data with good outputs. The usefulness of such bug detection/repair capabilities clearly extends to network configuration as well.

%that avoids  creating pairs of (bad, good) program examples for training using heuristics, which leads to non representative training data and models that do not generalize. The usefulness of such bug detection/repair capabilities clearly extends to  network configuration as well.

%\ZH{A similar use case has been tackled in networking by vanbever cite network auto complete..}
%\DR{rephrase, merge: }\ZH{todo} There are of course other approaches for bug detection following the same principles (e.g.\cite{pradel2018deepbugs,pradel2020neural}. We refer the reader to the nice related work section in the \href{https://arxiv.org/pdf/2011.07986.pdf}{Neural software analysis} paper, which, based on the use cases of bug detection, code completion and type prediction, also analyzed when and when not to use ML techniques to automate software analysis tasks.

\section{Transfer from programming to configuration languages: a reality check}\label{sec:transfer} 

In this section, we  summarize the literature we exposed previously from a different angle. In particular, we aim at summarizing the lessons learned from NLP applications to computer language tasks (in terms of challenges, model training requirements, data availability, data pre-processing, solution quality, etc.),  projecting the expected impact of applying these NLP techniques to network configuration tasks.

\begin{figure}[t]
\centering
    \includegraphics[width=0.65\columnwidth]{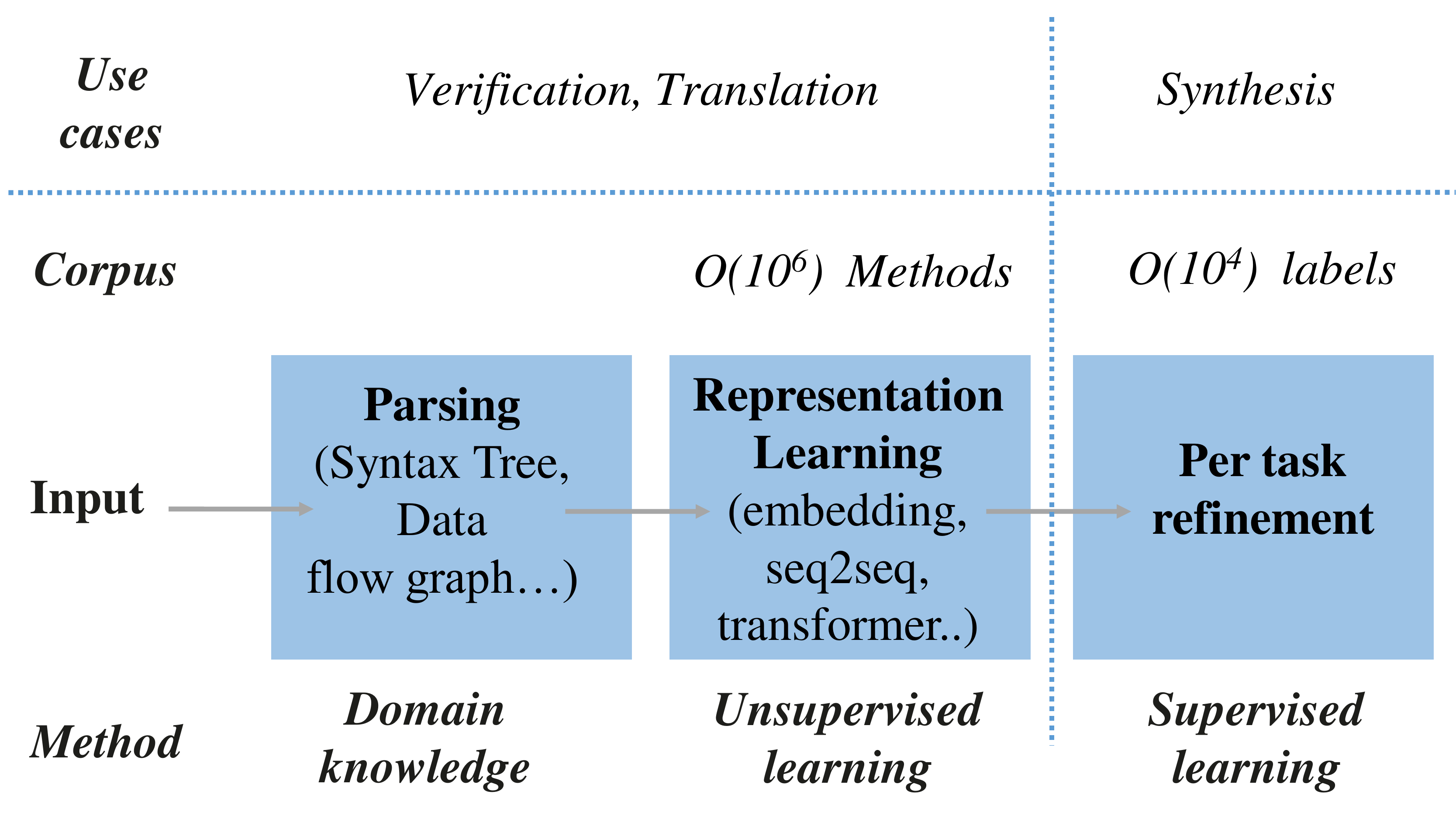} 
    \caption{Learning from programming languages: a generic pipeline}
    \label{fig:architecture}
\end{figure}

%%%%%%%%%%%%%%%%%%%%%%%%%%%%%%%%%%%%%%%%%%%%%%%%%
\subsection{Limits of current approaches} %%%%%%%%%%%%%%%%%%%%%%%%%%%%%%%%%%%%%%%%%%%%%%%%%%%%%%%%%%%%%%%%%%%%%%%%%%%%%%%
\paragraph{Lessons from computer language field} 
Despite the tremendous advances of NLP for  programming language tasks, limits still exist.
In recent work, \href{https://arxiv.org/pdf/2105.04297.pdf}{Peng~\textit{et al.}}\cite{peng2021could}  argue for the need to take more into account the specificities of programming languages, which are obviously different from natural language.
As we mentioned earlier, prior work used to counter this lack of semantic understanding by adding expert features such as data flow graph~\cite{li2019improving}. 
%Authors in \cite{peng2021could} propose to 
For instance, programming language theory can formally define the semantics of a program, seeing it as an entity that modifies its environment (e.g. memory) and performing elementary operations (e.g. I/O): this helps in \cite{peng2021could} to learn  \textit{intermediate code representations} that align well with operations defined in formal semantics and additionally leverage information on environment changes.  
Similarly, \href{https://arxiv.org/pdf/2009.07235.pdf}{Chakraborty~\textit{et al.}}~\cite{chakraborty2021deep} investigate the potential of deep learning methods in uncovering software bugs and vulnerabilities. Their careful analysis sheds light on poor performance that the authors attribute to a set of issues related to (i) unrealistic or bad-quality training data (ii) simplistic models (e.g. simple tokenization instead of taking into account code structure). Interestingly, their analysis reveals that whenever deep learning models perform well,  it may be due to the ``wrong reasons" (e.g. leveraging specific user-defined variable or function names to take their decisions, instead of more fundamental bug causes).
Last, as argued by \href{https://cacm.acm.org/magazines/2022/1/257443-the-growing-cost-of-deep-learning-for-source-code/fulltext}{Hellendoorn and Sawant}~\cite{hellendoorn2021growing}, a major limit of current approaches remains the exorbitant cost, both monetary and data-wise, to train the most performing models -- as we further develop in the next subsections.

\paragraph{Impact on network configuration}
Limits of NLP for computer programs are still far from being well understood. But whatever these limits are, the same problems can be expected to rise in network-use cases: for example, the difficulty of taking into account programming languages specificity, is likely to also apply to network configuration languages, which also have their own structured semantics. Unsurprisingly, the lack of large volumes of good quality  applies also to network use-cases, and so does the cost of training large models, for which the upfront cost to bootstrap NLP techniques for network configuration can be expected to be sizeable.
%\EX{Given all this, our network administrator For instance, concerning our BGP network admin }

\subsection{Model and training complexity}
\paragraph{Lessons from computer language field} 
We here report examples of model size and training time for computer language (whereas training cost for few models is extrapolated in~\cite{hellendoorn2021growing,lambdalabs}).  
Austin \textit{et al.}'s models~\cite{austin2021program} ranged between 244 million and 137 billion non-embedding parameters. 
CodeTrans\cite{elnaggar2021codetrans} starts with small models of 60 million parameters (17 days of training) but the large models have nearly 800 millions (nearly 90 days). Small models are trained on one single NVIDIA GPU Quadro RTX 8000, while larger models relied also on a number of Google TPUs (8$\times$ TPUv2 and 8$\times$ TPUv3).
PLBART~\cite{ahmad2021unified} employs 140M parameters and (trained in about 2 weeks on 8 Nvidia GeForce RTX 2080 GPU).
Codex~\cite{chen2021evaluating} models containing up to 12B parameters (costing few hundreds of petaflop/s-days). 
To give an idea, GPT-3~\cite{brown2020language}, one of the current largest natural language models  has 175 billion parameters (and would need 355 years of training on one Tesla V100 GPU). Thus, we see that  most programming-language models have a high complexity, comparable  to natural language ones: this is the case because such models take both natural language and code as input to training (e.g. text to code) and hence, they need both capabilities. 

\paragraph{Impact on network configuration}
Assuming the right data for training is available (which we assess next), the impact on network configurations depends mainly on the use case and whether it needs only configuration language or also natural language as input. Training both on natural language and configuration languages would lead to models at least as complex as natural languages ones, so that only a handful of actors could afford to train them. Therefore, for network configuration synthesis from high-level language, it is more reasonable to start from pre-trained models on natural language (as opposite to train from scratch) and fine tune them with hybrid natural and configuration language data.
Conversely, natural language is not needed for both network configuration verification and translation, where training from scratch may be more feasible (given limited vocabulary size and its complexity). 

\subsection{Data availability and corpus sizes}\label{sec:transfer:data}
\paragraph{Lessons from computer language field} 
Complementarily, we summarize the amounts of data exploited to train NLP models  for code-related tasks: in almost all existing literature~\cite{sawant2021naturally}, the (i) unsupervised pre-training phase uses huge amounts of data, while the (ii) supervised refinement needed reasonable amounts of labels. 
Examining the volumes for the (i) \emph{unsupervised pre-training phase}, for instance Code2vec~\cite{alon2019code2vec} used a dataset of 14\,M methods for unsupervised training. Li \textit{et al.}~\cite{li2019improving} similarly leveraged a dataset of almost 5\,M methods. Austin \textit{et al.}\cite{austin2021program} trained on almost 3\,B documents (web, dialog and wikipedia), which were tokenized into 3\,T byte-pair-encoding (BPE) tokens with a vocabulary of 32\,K tokens. Not all those documents however pertained to code: data including both code and text constituted ``only'' around 14\,M documents (roughly 18\,B BPE tokens).
Similarly, PLBART~\cite{ahmad2021unified} used 36\,B (Python), 28\,B (Java) and 7\,B (StackOverflow natural language) tokens for pre-training extracted from 224\,GB (Python),  352\,GB (Java) and 79\,GB (StackOverflow HTML)  code respectively.
Although fine-tuned from GPT-3 family models that have already learned  good NLP representations, Codex~\cite{chen2021evaluating} was further trained on 179\,GB of unique python files scraped from GitHub.

Examining the volumes for the (ii) \emph{supervised refinement phase}, 
CodeTrans\cite{elnaggar2021codetrans} needed 100\,K labeled samples for its various tasks (as opposed to nearly 8\,M samples, all programming languages included, for the unsupervised pre-training). For comparison, the NL2bash natural langage to bash dataset~\cite{lin2018nl2bash} used for the text to bash NeurIPS NLC2CMD Competition in 2020\cite{agarwal2021neurips} contained 10\,K pairs.

\paragraph{Impact on network configuration}
Overall, it is the online accessibility of a plethora of code-related data (source code, online questions and answers etc.) that fueled breakthrough and successful applications of NLP to programming languages~\cite{sawant2021naturally}.  This is clearly less the case for network configuration languages, where publicly available data is surely not abundant.  The three use cases we consider need raw network configurations for the unsupervised representation learning step: although corpus size requirements are much less stringent than those of natural language, they remain considerable. Additionally, unlike programming languages data which is abundant in various software repositories and community forums, real network configuration data is extremely scarce.  For instance, reconsidering datasets used in the network configuration literature (see Sec.~\ref{sec:formal}), the largest dataset is Bagpipe~\cite{weitz2016scalable} which was tested on 3 autonomous systems totalling just 240K lines of BGP router configurations. 

% Many of network configuration work starts, not from raw configurations in vendor language, but from abstractions thereof. 
As a consequence, only few big ISPs and vendors could afford in theory to create such datasets.
To offset this problem, pooling of datasets (e.g., from university IT) seems a way forward -- though this can open security risks. Another option is to synthetically generate network configurations for various topologies -- with the risk of oversimplification. Alternatively, some narrow use-cases may benefit from automatically collecting, at the same time, pairs of configurations and corresponding forwarding tables and states, which would allow training models to generate one from the other. Generally speaking, the lack of large volumes of good quality and real data is expected to be a more stringent limitation than the model complexity and training cost in our opinion.

\subsection{Input pre-processing}
\paragraph{Lessons from computer language field} 
Similarly to tokenization for natural language, programming languages, and network configuration languages alike, need some pre-processing beforehand. 
Many of the tools described earlier employ pre-processors to extract other intermediate representations such as Abstract Syntax Trees (AST) from source code before feeding them to Neural Networks. 
\href{https://tree-sitter.github.io/tree-sitter/}{tree-sitter} to extract AST from code. 
Further work has studied the importance of using additional abstractions  
\cite{allamanis2017learning,li2019improving} or learning more sophisticated  representations\cite{chirkova2021empirical}. 
Allamanis \textit{et al.}~\cite{allamanis2017learning} \href{https://arxiv.org/abs/1711.00740}{proposed} to first use graph structures to represent code (e.g. take into account long range dependencies between variables) then use gated graph neural networks to learn from code for toy example applications such as \textit{Varnaming} (finding the best variable name given its usage) and \textit{Varmisuse} (predicting which variable to use at which place). In addition to AST, {Li \textit{et al.}}~\cite{li2019improving} use also program dependence graph and data flow graph. Conversely,  Chirkova \textit{et al.}\cite{chirkova2021empirical} study   whether transformers architecture are good to process AST instead of raw code investigating several design choices -- testifying that this is still an open area for future research.  

%performed an \href{https://arxiv.org/pdf/2010.07987.pdf}{empirical analysis}, asking the question of whether transformers can successfully use code syntactic structure. As applications, they evaluated on the tasks of code completion, function naming and bug fixing. Their code is available \href{https://github.com/bayesgroup/code_transformers}{online}. Unlike other work which applies transformers on natural language describing code, they assess whether transformers are good to process AST instead. The paper compares multiple design choices. For example, the use of positional encoding versus more sophisticated relative attention to take into account order of elements in the sequence. They also compare two modes: full-data mode and anonymized mode in which user defined values are replaced by placeholders (meaning all variables are consistently renamed (e.g. var1, var2, etc.) throughout the code). The paper concludes that it's better to ensemble transformers that use both modes for better performance.

\paragraph{Impact on network configuration}
Clearly, appropriate parsers must be used for network configuration languages as well.  Batfish~\cite{fogel2015general} is one popular example to extract a network model from configurations, which could be leveraged to build the necessary input to train NLP network configuration language models. As Batfish limitedly support a number of cases/languages, it  could be complemented by  custom parsers developed as side-contributions in other work (for example, in their study of network configuration complexity, Benson \textit{et al.}\cite{benson2009unraveling} leverage the syntax contained in the documentation to manually create a grammar from router configurations).
Generally speaking, we make two observations. First, as stressed by \href{https://ieeexplore.ieee.org/abstract/document/4660333}{Caldwell \textit{et al.}}\cite{caldwell2008adaptive} the main challenge is tied to the parser maintenance cost, because of frequent changes to configuration languages and features.  Second, network configuration language are syntactically poor, as they contain a lot of  ``identifiers'' that only have local semantic (e.g., constant values, weights, IP addresses): while pre-processing may help letting the structure of the (relatively simple) syntax emerge, it may be more difficult to let the tacit structure of (significantly more complex) arbitrarily selected identifiers emerge.

\subsection{Expected performance}
\paragraph{Lessons from computer language field} 
Despite progress is rapidly made, with  models able to solve problems without even training on code-only datasets~\cite{austin2021program} and as others empower production-level tools~\cite{chen2021evaluating},
it is difficult to understand, already on the flourishing field of NLP for programming languages, up to which level nowadays models are mature for real deployment.
Indeed, with the exception of GitHub Copilot~\cite{chen2021evaluating}, most of the use cases are so far limited to toy examples with  limited complexity (e.g.,  method or variable naming).
Code verification and bug detection has been done mainly for simple tasks such as off-by-one bugs~\cite{briem2019using}.

To get a concrete sense of the complexity of the tasks that are solved by today models, the reader can refer to the largest available benchmarks against which the models are evaluated. 
For example, one of the largest evaluations of NLP for code related tasks~\cite{austin2021program} runs its models on two benchmarks that were made public: the Mostly Basic Programming Problems (\textit{MBPP}) dataset\footnote{\url{https://github.com/google-research/google-research/tree/master/mbpp}} and the \textit{MathQA}\footnote{\url{https://math-qa.github.io/}}.
Despite their relatively large number (974 and 23914 respectively), the problems remain fairly simple, barely matching the skills of entry-level programmers. 
Thus, existing models perform well but on rather easy tasks, and it thus unclear to project expected performance on more useful and realistic tasks.

%Another example that the reader can refer to is \href{https://arxiv.org/abs/2102.04664}{\textit{CodeXGLUE}}\cite{lu2021codexglue}\footnote{\url{https://github.com/microsoft/CodeXGLUE}}, a recent thorough benchmark for 10 tasks spanning 14 datasets, including the tasks of interest in this paper but also beyond with, e.g., code clone detection, defect detection, repair, documentation translation etc. 
%\ZH{say a word about how easy/complex the tasks are}
%{CodeXGLUE: A Machine Learning Benchmark Dataset for Code Understanding and Generation}

Furthermore, quality of the resulting code is also of not straightforward evaluation as stressed by  \href{https://arxiv.org/pdf/2012.07581.pdf}{Agarwal \textit{et al.}}\cite{agarwal2020quality} in the case of code translation. Similarly,  \href{https://arxiv.org/abs/2009.10297}{Codebleu}\cite{ren2020codebleu} proposes a metric to evaluate code synthesis tasks, that aims to be as close as possible to human ratings for the three tasks of text to code synthesis,  translation and  refinement.
Thus, more work as~\cite{agarwal2020quality,ren2020codebleu} seems needed to programmatically evaluate  generated software, especially when output of tasks will span thousands of lines per task, as opposed to few lines today.

%When talking about performance, however, another under-estimated issue is that evaluation of new solutions is not always straightforward neither. For example, \href{https://arxiv.org/pdf/2012.07581.pdf}{Agarwal \textit{et al.}}\cite{agarwal2020quality} stresses the problem of estimating the quality of the translated code.
%Similarly, \href{https://arxiv.org/abs/2009.10297}{Codebleu}\cite{ren2020codebleu} proposes a metric to evaluate code synthesis tasks that aims to be as close as possible to human ratings, in three code-related tasks: text to code synthesis, code translation and code refinement. More work seems needed to properly evaluate programmatically generated software (especially for future realistic tasks spanning thousands of lines per task, as opposed to few lines today).

\paragraph{Impact on network configuration}
In reason of the above limits, speculating  how much of this research would benefit realistic  network configuration use cases is not an easy task. Despite great progress on programming language, the performance is still limited for both synthesis and translation, especially taking into account how elementary the tasks are.

From a practical, cultural and historical perspective, networks strives to achieve ``four nines'' to ``five nines'' reliability~\cite{fivenines}.
Under this perspective, even an almost perfect ``four nines'' ML model (resp. five nines), i.e., a model that is correct 99.99\% (resp. 99.999\%) of the times, would still generate 10 (resp. 1) configuration errors per 100k configuration lines. Otherwise stated, an almost perfect network configuration obtained via ML synthesis or translation, remains an incorrect configuration: no rational operator would accept to use it as-is. Additionally, finding ML-generated configuration bugs can be more difficult for humans than finding their own bugs.

Of course, proper verification and correction tools could be developed in parallel. Besides, as rightly argued by  \href{https://dl.acm.org/doi/abs/10.1145/3397481.3450656}{Weisz et al.}\cite{weisz2021perfection}, ``perfection is may be not required'' as humans and AI could ``partner" for code translation and other tasks. This was empirically shown as well by Austin et al.~\cite{austin2021program} who demonstrated that chatting with the model in natural language could help in practice improving the performance on the task.  
The ML model could be thus an assistant that eases the work of the engineer in generating the first translation or code, whereas the latter could build on it to generate the final configuration. For this to happen, proper tools are required to inspect, explain and debug the work of ML models.

%\cite{pradel2020neural}\href{https://arxiv.org/pdf/2011.07986.pdf}{Neural software analysis}: bug detection, code completion and type prediction and analyzes when and when not to use software analysis.

% \begin{table*}[]
% \begin{tabular}{l|l|l|l|l|l}
%       & Labels  & Availability & Unsupervised & Corpus & Pre-processing \\ \hline
% Synthesis    & Spec-to-conf & Challenging    & Not possible &    ~100K to ~1B         &   Doable            \\ \hline
% Verification & \begin{tabular}[c]{@{}l@{}} Not manadatory \\ correct/incorrect confs \end{tabular} & Possible     & Possible    &     ~100K to ~1B    &      Doable         \\ \hline
% Translation  &  Not manadatory  & Possible & Possible &   ~100K to ~1B  &       Doable      &              
% \end{tabular}\caption{summary of requirements of each use case: all use cases need large amounts of raw configuration data for pretraining/unsupervised learning.}\label{summary:tab}
% \end{table*}
\begin{table*}[t]
\footnotesize
\caption{Summary of data requirements for the different network configuration use-cases}\label{summary:tab}
\begin{tabular}{lp{3.5cm}lp{5.0cm}}
\toprule 
{\bf Use-case} & {\bf Data and labels}   & {\bf Availability }  &  {\bf Difficulty/comment } \\ 
\midrule
\multirow{2}{*}{Synthesis} &  Spec-to-conf &  Possible~\cite{birkner2020config2spec} & Almost ready for trial\\ 
                          & Text-to-conf  &   Difficult~\cite{birkner2018net2text} & Very hard at present stage \\
\cmidrule(r){2-4}
\multirow{3}{*}{Verification} &  Binary (``bogus'' or not) Segmented (``bogus'' parts  & Possible & Easy for synthetic errors~\cite{yasunaga2021break,richter2021deepmutants}, harder for real errors. Unsupervised verification does not need labels \\
     % & Segmented (``bogus'' parts)   & Possible  \\
     %&  Unsupervised verification & possible without labels \\  

\cmidrule(r){2-4}
Translation  &  Mono-language corpora  &  Difficult & 
    Need for multiple datasets  \\

\bottomrule
\end{tabular}
\end{table*}

\section{Conclusion and recommendations}\label{sec:conclusion}
In this paper, we overview recent progress on NLP application to computer languages,  and project its potential impact to the area of automated network configuration. In light of our analysis, we make the following conclusive observations:

\begin{itemize}
    \item Generally speaking, the \emph{bootstrap cost} in terms of amounts of data, as well as computational power for unsupervised learning, could be affordable but only for top largest vendors and ISPs.  Especially, we assess that \emph{data cost} primes over the model training cost:  unlike the computer language field, where large corpora are available over the Internet, network configuration examples are scarce.   That is to say, academic community may have a high entry barrier unless a community-wide effort is made a priori to make a systematic, organized and organic collection. 
    
    \item Additionally, unlike computer programming language that are semantically rich, configuration languages are likely to contain a large amount of identifiers with local significance (e.g., addresses, interfaces, tunnels, etc.) for which the importance of structured network \emph{topological metadata} is likely to be of key importance (similarly to AST, PDG, DFG graph structures for programming language).
 \end{itemize}

\noindent We additionally report
in Table~\ref{summary:tab}  the requirements for each use case, that we summarize as follows:

  \begin{itemize}
    \item For what concerns \emph{configuration verification}, we estimate that anomaly detection on top of learned configuration language models has realistic potential for application. Indeed,
    provided that few false positive are generated, even if the error recall rate is not perfect,  automated detection of  configuration errors is helpful and may have a readily practical impact.

    \item The expected performance for code/configuration synthesis is still elementary given the use cases tried so far are rather simple. As such \emph{configuration synthesis} is for the time being still a ``moonshot'', especially that the text/specs from which to generate the configuration is a large challenge. A second, non lesser, challenge is reliability: one cannot push to production networks a configuration that might contain some bugs (a 99.9\% correct model still generates 100 bugs for 100K lines configuration) so that extremely reliable verification is a  mandatory pre-condition for significant development in this area.
        %\DR{ So unless coupled with reliable verification, it's going to be tough... But from discussions during CoNext, it's just a matter of very short time before first papers on this will appear.}
        
    \item 
    The same applies but to a lesser extent for \emph{configuration translation}: current performance of software/code translation between programming languages does not need labels and is very promising, but the fact that it might contain translation errors, might make it hard to use in practice today. We expect this use case to be further developed, conditioned to the success of the previous two.
 
\end{itemize}

More broadly, we point out that, beyond the network configuration use cases discussed in this paper, several networking problems~\cite{CoNEXT21-b,ring2017ip2vec,cohen2020dante,velickovic2018graph,kool2018attention} are already benefiting from NLP techniques. As such, a growing research trend is expected to emerge in the community, increasing the availability, spread and knowledge of NLP tools for networking in general.  Overall, we believe that given the amount of work that has been done on programming languages, and given the reach of NLP to other networking use-cases already,  it is only a matter of time before these techniques are actively researched, and successfully applied, onto network configuration languages.

\EX{At the same time, we want to express a word of caution --- while network languages will largely benefit from progress on NLP for programming languages, the two fields are largely different.  For the computer programming languages, some argue that  ``Automated AI tools will work hand in hand with software developers. The focus of software engineering will move from writing code from scratch to reviewing code written and tested by AI.'' (Thomas Zimmermann)~\cite{zimmermann}.  For the network configuration case, as long as NLP tools are not perfect, manual checking of large volumes of automatically generated configuration to spot rare yet critical errors generated by NLP tools seems less appealing. Having this difference clear in mind, can help devising a successful research agenda.}

% \begin{appendices}

% \section{USeful info}

% \end{appendices}
\bibliographystyle{elsarticle-num} 
\bibliography{transconfiguration}

\begin{thebibliography}{10}
\expandafter\ifx\csname url\endcsname\relax
  \def\url#1{\texttt{#1}}\fi
\expandafter\ifx\csname urlprefix\endcsname\relax\def\urlprefix{URL }\fi
\expandafter\ifx\csname href\endcsname\relax
  \def\href#1#2{#2} \def\path#1{#1}\fi

\bibitem{fogel2015general}
A.~Fogel, S.~Fung, L.~Pedrosa, M.~Walraed-Sullivan, R.~Govindan, R.~Mahajan,
  T.~Millstein, A general approach to network configuration analysis, in: 12th
  USENIX Symposium on Networked Systems Design and Implementation (NSDI 15),
  2015, pp. 469--483.

\bibitem{xkcd927}
\url{https://xkcd.com/927/}.

\bibitem{lample2017unsupervised}
G.~Lample, A.~Conneau, L.~Denoyer, M.~Ranzato, Unsupervised machine translation
  using monolingual corpora only, arXiv preprint arXiv:1711.00043 (2017).

\bibitem{brown2020language}
T.~Brown, B.~Mann, N.~Ryder, M.~Subbiah, J.~D. Kaplan, P.~Dhariwal,
  A.~Neelakantan, P.~Shyam, G.~Sastry, A.~Askell, et~al., Language models are
  few-shot learners, Advances in neural information processing systems
  (NeurIPS) 33 (2020) 1877--1901.

\bibitem{gpt3}
\url{https://openai.com/blog/gpt-3-apps/}.

\bibitem{CoNEXT21-b}
L.~Gioacchini, L.~Vassio, M.~Mellia, I.~Drago, Z.~Ben~Houidi, D.~Rossi,
  Darkvec: Automatic analysis of darknet traffic with word embeddings, in: ACM
  CoNEXT, 2021.

\bibitem{ring2017ip2vec}
M.~Ring, A.~Dallmann, D.~Landes, A.~Hotho, Ip2vec: Learning similarities
  between {IP} addresses, in: IEEE International Conference on Data Mining
  Workshops (ICDMW), IEEE, 2017, pp. 657--666.

\bibitem{cohen2020dante}
D.~Cohen, Y.~Mirsky, M.~Kamp, T.~Martin, Y.~Elovici, R.~Puzis, A.~Shabtai,
  Dante: A framework for mining and monitoring darknet traffic, in: European
  Symposium on Research in Computer Security, Springer, 2020, pp. 88--109.

\bibitem{velickovic2018graph}
P.~Veličković, G.~Cucurull, A.~Casanova, A.~Romero, P.~Liò, Y.~Bengio,
  \href{https://openreview.net/forum?id=rJXMpikCZ}{Graph attention networks},
  in: International Conference on Learning Representations, 2018.
\newline\urlprefix\url{https://openreview.net/forum?id=rJXMpikCZ}

\bibitem{kool2018attention}
W.~Kool, H.~van Hoof, M.~Welling, Attention, learn to solve routing problems!,
  arXiv:1803.08475 (2018).

\bibitem{prabhu2020plankton}
S.~Prabhu, K.~Y. Chou, A.~Kheradmand, B.~Godfrey, M.~Caesar, Plankton: Scalable
  network configuration verification through model checking, in: 17th USENIX
  Symposium on Networked Systems Design and Implementation (NSDI 20), 2020, pp.
  953--967.

\bibitem{weitz2016scalable}
K.~Weitz, D.~Woos, E.~Torlak, M.~D. Ernst, A.~Krishnamurthy, Z.~Tatlock,
  Scalable verification of border gateway protocol configurations with an smt
  solver, in: Proceedings of the 2016 acm sigplan international conference on
  object-oriented programming, systems, languages, and applications, 2016, pp.
  765--780.

\bibitem{beckett2017general}
R.~Beckett, A.~Gupta, R.~Mahajan, D.~Walker, A general approach to network
  configuration verification, in: Proceedings of the Conference of the ACM
  Special Interest Group on Data Communication, 2017, pp. 155--168.

\bibitem{tian2019safely}
B.~Tian, X.~Zhang, E.~Zhai, H.~H. Liu, Q.~Ye, C.~Wang, X.~Wu, Z.~Ji, Y.~Sang,
  M.~Zhang, et~al., Safely and automatically updating in-network acl
  configurations with intent language, in: ACM SIGCOMM, 2019, pp. 214--226.

\bibitem{gember2016fast}
A.~Gember-Jacobson, R.~Viswanathan, A.~Akella, R.~Mahajan, Fast control plane
  analysis using an abstract representation, in: Proceedings of the 2016 ACM
  SIGCOMM Conference, 2016, pp. 300--313.

\bibitem{feldmann1999netdb}
A.~Feldmann, Netdb: Ip network configuration debugger/database", in: AT\&T
  Software Symposium, Citeseer, 1999.

\bibitem{steffen2020probabilistic}
S.~Steffen, T.~Gehr, P.~Tsankov, L.~Vanbever, M.~Vechev, Probabilistic
  verification of network configurations, in: Proceedings of the Annual
  conference of the ACM Special Interest Group on Data Communication on the
  applications, technologies, architectures, and protocols for computer
  communication, 2020, pp. 750--764.

\bibitem{abhashkumar2020tiramisu}
A.~Abhashkumar, A.~Gember-Jacobson, A.~Akella, Tiramisu: Fast multilayer
  network verification, in: 17th USENIX Symposium on Networked Systems Design
  and Implementation (NSDI 20), 2020, pp. 201--219.

\bibitem{beckett2017network}
R.~Beckett, R.~Mahajan, T.~Millstein, J.~Padhye, D.~Walker, Network
  configuration synthesis with abstract topologies, in: Proceedings of the 38th
  ACM SIGPLAN Conference on Programming Language Design and Implementation,
  2017, pp. 437--451.

\bibitem{el2018netcomplete}
A.~El-Hassany, P.~Tsankov, L.~Vanbever, M.~Vechev, $\{$NetComplete$\}$:
  Practical $\{$Network-Wide$\}$ configuration synthesis with autocompletion,
  in: 15th USENIX Symposium on Networked Systems Design and Implementation
  (NSDI 18), 2018, pp. 579--594.

\bibitem{el2017network}
A.~El-Hassany, P.~Tsankov, L.~Vanbever, M.~Vechev, Network-wide configuration
  synthesis, in: International Conference on Computer Aided Verification,
  Springer, 2017, pp. 261--281.

\bibitem{liu2018automatic}
H.~H. Liu, X.~Wu, W.~Zhou, W.~Chen, T.~Wang, H.~Xu, L.~Zhou, Q.~Ma, M.~Zhang,
  Automatic life cycle management of network configurations, in: Proceedings of
  the Afternoon Workshop on Self-Driving Networks, 2018, pp. 29--35.

\bibitem{birkner2018net2text}
R.~Birkner, D.~Drachsler-Cohen, L.~Vanbever, M.~Vechev,
  $\{$Net2Text$\}$:$\{$Query-Guided$\}$ summarization of network forwarding
  behaviors, in: 15th USENIX Symposium on Networked Systems Design and
  Implementation (NSDI 18), 2018, pp. 609--623.

\bibitem{birkner2020config2spec}
R.~Birkner, D.~Drachsler-Cohen, L.~Vanbever, M.~Vechev, $\{$Config2Spec$\}$:
  Mining network specifications from network configurations, in: 17th USENIX
  Symposium on Networked Systems Design and Implementation (NSDI 20), 2020, pp.
  969--984.

\bibitem{gao2001stable}
L.~Gao, J.~Rexford, Stable internet routing without global coordination,
  IEEE/ACM Transactions on networking 9~(6) (2001) 681--692.

\bibitem{yen2021semi}
J.~Yen, T.~L{\'e}vai, Q.~Ye, X.~Ren, R.~Govindan, B.~Raghavan, Semi-automated
  protocol disambiguation and code generation, in: Proceedings of the 2021 ACM
  SIGCOMM 2021 Conference, 2021, pp. 272--286.

\bibitem{houidi2016knowledge}
Z.~B. Houidi, A knowledge-based systems approach to reason about networking,
  in: Proceedings of the 15th ACM Workshop on Hot Topics in Networks, 2016, pp.
  22--28.

\bibitem{benson2009unraveling}
T.~Benson, A.~Akella, D.~A. Maltz, Unraveling the complexity of network
  management., in: NSDI, 2009, pp. 335--348.

\bibitem{kim2011evolution}
H.~Kim, T.~Benson, A.~Akella, N.~Feamster, The evolution of network
  configuration: A tale of two campuses, in: Proceedings of the 2011 ACM
  SIGCOMM conference on Internet measurement conference, 2011, pp. 499--514.

\bibitem{lee2008automate}
S.~Lee, T.~Wong, H.~S. Kim, To automate or not to automate: on the complexity
  of network configuration, in: 2008 IEEE International Conference on
  Communications, IEEE, 2008, pp. 5726--5731.

\bibitem{habib2019neural}
A.~Habib, M.~Pradel, Neural bug finding: A study of opportunities and
  challenges, arXiv preprint arXiv:1906.00307 (2019).

\bibitem{richter2021deepmutants}
C.~Richter, H.~Wehrheim, Deepmutants: Training neural bug detectors with
  contextual mutations, arXiv preprint arXiv:2107.06657 (2021).

\bibitem{tian2020rethinking}
Y.~Tian, Y.~Wang, D.~Krishnan, J.~B. Tenenbaum, P.~Isola, Rethinking few-shot
  image classification: a good embedding is all you need?, in: Computer
  Vision--ECCV 2020: 16th European Conference, Glasgow, UK, August 23--28,
  2020, Proceedings, Part XIV 16, Springer, 2020, pp. 266--282.

\bibitem{alon2019code2vec}
U.~Alon, M.~Zilberstein, O.~Levy, E.~Yahav, code2vec: Learning distributed
  representations of code, Proceedings of the ACM on Programming Languages
  3~(POPL) (2019) 1--29.

\bibitem{briem2019using}
J.~A. Briem, J.~Smit, H.~Sellik, P.~Rapoport, Using distributed representation
  of code for bug detection, arXiv preprint arXiv:1911.12863 (2019).

\bibitem{li2019improving}
Y.~Li, S.~Wang, T.~N. Nguyen, S.~Van~Nguyen, Improving bug detection via
  context-based code representation learning and attention-based neural
  networks, Proceedings of the ACM on Programming Languages 3~(OOPSLA) (2019)
  1--30.

\bibitem{mikolov2013efficient}
T.~Mikolov, K.~Chen, G.~Corrado, J.~Dean, Efficient estimation of word
  representations in vector space, arXiv preprint arXiv:1301.3781 (2013).

\bibitem{grover2016node2vec}
A.~Grover, J.~Leskovec, node2vec: Scalable feature learning for networks, in:
  Proceedings of the 22nd ACM SIGKDD international conference on Knowledge
  discovery and data mining, 2016, pp. 855--864.

\bibitem{agarwal2021neurips}
M.~Agarwal, T.~Chakraborti, Q.~Fu, D.~Gros, X.~V. Lin, J.~Maene,
  K.~Talamadupula, Z.~Teng, J.~White, Neurips 2020 nlc2cmd competition:
  Translating natural language to bash commands, arXiv preprint
  arXiv:2103.02523 (2021).

\bibitem{ahmad2021unified}
W.~U. Ahmad, S.~Chakraborty, B.~Ray, K.-W. Chang, Unified pre-training for
  program understanding and generation, arXiv preprint arXiv:2103.06333 (2021).

\bibitem{chen2021evaluating}
M.~Chen, J.~Tworek, H.~Jun, Q.~Yuan, H.~Ponde, J.~Kaplan, H.~Edwards, Y.~Burda,
  N.~Joseph, G.~Brockman, et~al., Evaluating large language models trained on
  code, arXiv preprint arXiv:2107.03374 (2021).

\bibitem{austin2021program}
J.~Austin, A.~Odena, M.~Nye, M.~Bosma, H.~Michalewski, D.~Dohan, E.~Jiang,
  C.~Cai, M.~Terry, Q.~Le, et~al., Program synthesis with large language
  models, arXiv preprint arXiv:2108.07732 (2021).

\bibitem{yen2021tools}
J.~Yen, R.~Govindan, B.~Raghavan, Tools for disambiguating rfcs, in:
  Proceedings of the Applied Networking Research Workshop, 2021, pp. 85--91.

\bibitem{chen2018tree}
X.~Chen, C.~Liu, D.~Song, Tree-to-tree neural networks for program translation,
  arXiv preprint arXiv:1802.03691 (2018).

\bibitem{roziere2020unsupervised}
B.~Roziere, M.-A. Lachaux, L.~Chanussot, G.~Lample, Unsupervised translation of
  programming languages, Advances in Neural Information Processing Systems 33
  (2020).

\bibitem{roziere2021leveraging}
B.~Roziere, J.~M. Zhang, F.~Charton, M.~Harman, G.~Synnaeve, G.~Lample,
  Leveraging automated unit tests for unsupervised code translation, arXiv
  preprint arXiv:2110.06773 (2021).

\bibitem{roziere2021dobf}
B.~Roziere, M.-A. Lachaux, M.~Szafraniec, G.~Lample, Dobf: A deobfuscation
  pre-training objective for programming languages, arXiv preprint
  arXiv:2102.07492 (2021).

\bibitem{elnaggar2021codetrans}
A.~Elnaggar, W.~Ding, L.~Jones, T.~Gibbs, T.~Feher, C.~Angerer, S.~Severini,
  F.~Matthes, B.~Rost, Codetrans: Towards cracking the language of silicon's
  code through self-supervised deep learning and high performance computing,
  arXiv preprint arXiv:2104.02443 (2021).

\bibitem{pradel2020neural}
M.~Pradel, S.~Chandra, Neural software analysis, arXiv preprint
  arXiv:2011.07986 (2020).

\bibitem{yasunaga2021break}
M.~Yasunaga, P.~Liang, Break-it-fix-it: Unsupervised learning for program
  repair, arXiv preprint arXiv:2106.06600 (2021).

\bibitem{peng2021could}
D.~Peng, S.~Zheng, Y.~Li, G.~Ke, D.~He, T.-Y. Liu, How could neural networks
  understand programs?, arXiv preprint arXiv:2105.04297 (2021).

\bibitem{chakraborty2021deep}
S.~Chakraborty, R.~Krishna, Y.~Ding, B.~Ray, Deep learning based vulnerability
  detection: Are we there yet, IEEE Transactions on Software Engineering
  (2021).

\bibitem{hellendoorn2021growing}
V.~J. Hellendoorn, A.~A. Sawant, The growing cost of deep learning for source
  code, Communications of the ACM 65~(1) (2021) 31--33.

\bibitem{lambdalabs}
C.~Li, Lambda labs, \url{https://lambdalabs.com/blog/demystifying-gpt-3/},
  [Online; accessed 05-April-2022] (June 2020).

\bibitem{sawant2021naturally}
A.~A. Sawant, P.~Devanbu, Naturally!: How breakthroughs in natural language
  processing can dramatically help developers, IEEE Software 38~(5) (2021)
  118--123.

\bibitem{lin2018nl2bash}
X.~V. Lin, C.~Wang, L.~Zettlemoyer, M.~D. Ernst, Nl2bash: A corpus and semantic
  parser for natural language interface to the linux operating system, arXiv
  preprint arXiv:1802.08979 (2018).

\bibitem{allamanis2017learning}
M.~Allamanis, M.~Brockschmidt, M.~Khademi, Learning to represent programs with
  graphs, arXiv preprint arXiv:1711.00740 (2017).

\bibitem{chirkova2021empirical}
N.~Chirkova, S.~Troshin, Empirical study of transformers for source code, in:
  Proceedings of the 29th ACM Joint Meeting on European Software Engineering
  Conference and Symposium on the Foundations of Software Engineering, 2021,
  pp. 703--715.

\bibitem{caldwell2008adaptive}
D.~Caldwell, S.~Lee, Y.~Mandelbaum, Adaptive parsing of router configuration
  languages, in: 2008 IEEE Internet Network Management Workshop (INM), IEEE,
  2008, pp. 1--6.

\bibitem{agarwal2020quality}
M.~Agarwal, K.~Talamadupula, S.~Houde, F.~Martinez, M.~Muller, J.~Richards,
  S.~Ross, J.~D. Weisz, Quality estimation \& interpretability for code
  translation, arXiv preprint arXiv:2012.07581 (2020).

\bibitem{ren2020codebleu}
S.~Ren, D.~Guo, S.~Lu, L.~Zhou, S.~Liu, D.~Tang, N.~Sundaresan, M.~Zhou,
  A.~Blanco, S.~Ma, Codebleu: a method for automatic evaluation of code
  synthesis, arXiv preprint arXiv:2009.10297 (2020).

\bibitem{fivenines}
B.~Treynor, M.~Dahlin, V.~Rau, B.~Beyer, The calculus of service availability,
  Commun. ACM 60~(9) (2017) 42–47.
\newblock \href {https://doi.org/10.1145/3080202} {\path{doi:10.1145/3080202}}.

\bibitem{weisz2021perfection}
J.~D. Weisz, M.~Muller, S.~Houde, J.~Richards, S.~I. Ross, F.~Martinez,
  M.~Agarwal, K.~Talamadupula, Perfection not required? human-ai partnerships
  in code translation, in: 26th International Conference on Intelligent User
  Interfaces, 2021, pp. 402--412.

\bibitem{zimmermann}
\url{https://www.acm.org/articles/people-of-acm/2022/thomas-zimmermann}.

\end{thebibliography}

\end{document}